\documentclass[aps,twocolumn,pra,superscriptaddress,amsmath,tightenlines]{revtex4}
\usepackage{epsfig,graphicx,times}
\usepackage{amstext}
\usepackage{amsmath}            
\usepackage{amssymb}            
\usepackage{graphicx}           
\usepackage{latexsym}
\usepackage{color}
\usepackage{bm}
\begin{document}
\title{An efficient and compact quantum switch for quantum circuits}
\author{Yulin Wu}
\affiliation{Institute of Physics and Beijing National Laboratory for Condensed Matter Physics, Chinese Academy of Sciences, Beijing 100190, China}
\author{Li-Ping Yang}
\affiliation{Beijing Computational Science Research Center, Beijing 100193, China}
\author{Yarui Zheng}
\affiliation{Institute of Physics and Beijing National Laboratory for Condensed Matter Physics, Chinese Academy of Sciences, Beijing 100190, China}
\author{Hui Deng}
\affiliation{Institute of Physics and Beijing National Laboratory for Condensed Matter Physics, Chinese Academy of Sciences, Beijing 100190, China}
\author{Zhiguang Yan}
\affiliation{Institute of Physics and Beijing National Laboratory for Condensed Matter Physics, Chinese Academy of Sciences, Beijing 100190, China}
\author{Yanjun Zhao}
\affiliation{Institute of Microelectronics, Tsinghua University,
Beijing 100084, China}
\author{Keqiang Huang}
\affiliation{Institute of Physics and Beijing National Laboratory for Condensed Matter Physics, Chinese Academy of Sciences, Beijing 100190, China}
\author{William J. Munro}
\affiliation{NTT Basic Research Laboratories, NTT Corporation, 3-1 Morinosato-Wakamiya,
Atsugi, Kanagawa, 243-0198, Japan}
\author{Kae Nemoto}
\affiliation{National Institute of Informatics, 2-1-2 Hitotsubashi, Chiyoda-ku, Tokyo 101-8430,
Japan}
\author{Dong-Ning Zheng}
\affiliation{Institute of Physics and Beijing National Laboratory for Condensed Matter Physics, Chinese Academy of Sciences, Beijing 100190, China}
\author{C. P. Sun}
\affiliation{Beijing Computational Science Research Center, Beijing 100193, China}
\affiliation{Synergetic Innovation Center of Quantum Information and Quantum Physics,
University of Science and Technology of China, Hefei, Anhui 230026, China}
\author{Yu-xi Liu}
\email{yuxiliu@mail.tsinghua.edu.cn}
\affiliation{Institute of Microelectronics, Tsinghua University,
Beijing 100084, China}
\affiliation{Tsinghua National Laboratory for Information Science and Technology (TNList), Beijing 100084, China}
\author{Xiaobo Zhu}
\email{xbzhu16@ustc.edu.cn}
\affiliation{Institute of Physics and Beijing National Laboratory for Condensed Matter Physics, Chinese Academy of Sciences, Beijing 100190, China}
\affiliation{Synergetic Innovation Center of Quantum Information and Quantum Physics,
University of Science and Technology of China, Hefei, Anhui 230026, China}
\author{Li Lu}
\affiliation{Institute of Physics and Beijing National Laboratory for Condensed Matter Physics, Chinese Academy of Sciences, Beijing 100190, China}

\date{\today}

\maketitle
\textbf{The engineering of quantum devices has reached the stage where we now have small scale quantum processors containing multiple interacting qubits within them. Simple quantum circuits have been demonstrated and scaling up to larger numbers is underway~\cite{Nature464,book}. However as the number of qubits in these processors increases, it becomes challenging to implement switchable or tunable coherent coupling among them. The typical approach has been to detune each qubit from others or the quantum bus it connected to~\cite{Nature464,book}, but as the number of qubits increases this becomes problematic to achieve in practice due to frequency crowding issues. Here, we demonstrate that by applying a fast longitudinal control field to the target qubit, we can turn off its couplings to other qubits or buses (in principle on/off ratio higher than 100 dB). This has important implementations in superconducting circuits as it means we can keep the qubits at their optimal points, where the coherence properties are greatest, during coupling/decoupling processing. Our approach suggests a new way to control coupling among qubits and data buses that can be naturally scaled up to large quantum processors without the need for auxiliary circuits and yet be free of the frequency crowding problems.}

\begin{figure}[h]
\centering
\includegraphics[scale=0.65]{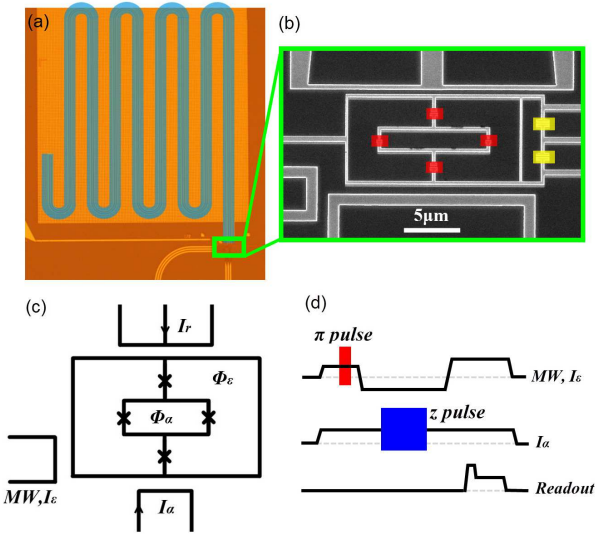}
\caption{{\bf Sample and measurement scheme.} (a) Optical micrograph of the flux qubit-resonator sample, the coplanar wave guide resonator is marked out by a blue ribbon. (b) SEM image of the small structures of the flux qubit circuit, Josephson junctions of the qubit and the readout dc-SQUID are marked out by red and yellow boxes, respectively. (c) Schematic of the gap tunable flux qubit with control and coupling lines. The smaller junction of the three-junction flux qubit is replaced by a dc-SQUID, called the $\alpha$-loop, in which   the flux $\Phi_{\alpha}$ threading the $\alpha$-loop is tuned by the current $I_{\alpha}$ through the ${\alpha}$ bias line. The ${\alpha}$ bias line is also used for applying longitudinal control pulses($z$ pulses) to perform quantum switch. The qubit is coupled to the resonator  through the mutual inductance between the qubit loop and an antenna which shorts the resonator at one end. Qubit flux bias $\Phi_{\epsilon}$ and microwave pulses are generated by the current $I_{\epsilon}$  through another  microwave line on the left. To achieve high flux bias stability, the flux qubit adopts a gradiometeric geometry.  $I_{r}$ denotes the zero-point current of the resonator. The readout dc-SQUID is not shown here. (d) A schematic of the qubit control and readout pulsing sequences.The qubit is prepared at the excited state by applying a microwave $\pi$ pulse, away from the optimal point where the qubit is largely detuned from the resonator, then the qubit is shifted to the coherent optimal point to be coupled with the resonator. While the qubit and the resonator are coupled together, longitudinal control pulses($z$ pulses) are applied by the ${\alpha}$ bias line to switch on/off the coupling {\itshape in-situ}. After all operations are done, the qubit is again shfited away from the optimal point and the qubit state is read out by pulsing the readout dc-SQUID and detecting its switching. }\label{fig1}
\end{figure}

Superconducting quantum circuits~\cite{r1,r2} are promising candidates to realize quantum processors and simulators. They have been used to demonstrate various quantum algorithms and implement thousands of quantum operations within their coherence time~\cite{Martinis-2015}, in which controllable couplings are inevitable.  The typical way to couple/decouple two superconducting quantum elements with always-on coupling is to tune their frequencies in or out of resonance~\cite{DT1,DT2,DT3,DT4,DT5,DT6,DT7}. This method is widely adopted even for the most recent universal gate implementations~\cite{Martinis-2015,G1,G2,G3}. However, it suffers from several defects, namely, it is technically difficult to avoid frequency crowding problem in large scale circuits; the qubits cannot always work at the coherent optimal point and the fast tuning of the qubit frequency results in non-adiabatic information leakage. To overcome the above problems, significant effort has been devoted both theoretically ~\cite{CT1,CT2} and experimentally ~\cite{CE1,CE2,CE3,CE4,CE5,CE6} to develop couplers for parametrically tuning the coupling strength between two elements. Recently high coherence and fast tunable coupling has been demonstrated~\cite{Yu-PRL-2014}, however, these quantum or classical couplers increase the complexity of the circuits and introduce new decoherence sources. Therefore, the implementation of quantum switch for coherent coupling between quantum elements is still a big challenge in scalable quantum circuits.

\begin{figure}[h]
\centering
\includegraphics[width=8.5 cm]{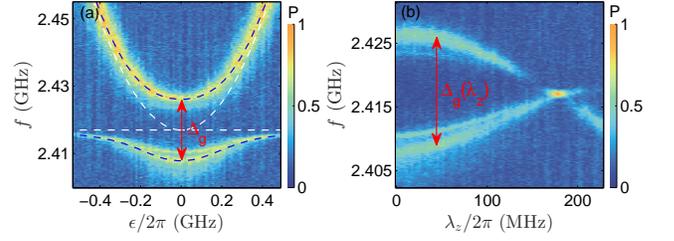}
\caption{ {\bf Spectrum of the qubit-resonator.} (a) Spectrum of the gap tunable flux qubit coupled to the resonator as a function of $\varepsilon$. The qubit is driven by a microwave pulse with varied frequency $f$, when $f$ matches the qubit frequency $\omega_{qb}$, the qubit is excited to the exited state, results in a increase of the excited state occupation probability P. The frequency $\omega_r/2\pi$ of the resonator is 2.417 GHz, indicated by the horizontal white dashed line; The other white dashed line shows the qubit frequencies at zero coupling strength. Blue dashed curves are fit of the spectrum lines. An anti-crossing gap $\Delta_{g}=2g/2\pi=18.28$ MHz (red line with two arrows) of the spectrum lines due to the coupling is clearly visible. (b) Under longitudinal control, the anti-crossing gap $\Delta_{\rm g}(\lambda_{z})$ (red line with two arrows) variation with the amplitude $\lambda_z$ of the control pulse at the qubit optimal point. $\Delta_{\rm g}$ decreases when increasing $\lambda_z$ from zero, and reaches an invisible minimum value at the {\itshape switch-off point} $\lambda_z = \lambda_{\rm zoff}\approx 1.2\,\omega_{z}$.  Here, the frequency corresponding to the {\it switch-off point} is $\omega_z/2\pi=150$ MHz. Color indicates the normalized occupation probability P of the qubit excited state in both (a) and (b).}\label{fig2}
\end{figure}

In this letter, we demonstrate a simple yet reliable method to switch on/off the coupling between a quantum resonator and a superconducting flux qubit~\cite{F1} via a control field, longitudinally applied to the qubit~\cite{ZLPRA,YXLNJP}.
Our system is a gap tunable flux qubit~\cite{gap1,gap2}, coupled to a $\lambda/4$ coplanar wave guide  resonator through mutual inductance ($M=0.74$ pH). The switching pulse is applied to the qubit through the qubit $\alpha$-loop (see Methods). When there is no switching pulse, the qubit-resonator Hamiltonian can be written as~\cite{MooijNature04}
\begin{equation}
H_{\rm QR}=\frac{\hbar}{2}(\Delta\sigma_{z}+\varepsilon\sigma_{x})+\hbar\omega_{r} a^{\dagger}a+\hbar g( a^{\dagger}+a)\sigma_{x}, \label{eq:1}
\end{equation}
here, $\hbar\Delta$ is the energy gap of the qubit, $\hbar\varepsilon=2I_{p}(\Phi_{\epsilon}-0.5\Phi_{0})$ is the energy difference between the two persistent current states, $\Phi_{\epsilon}$ is the magnetic flux through the qubit loop, $I_{p}$ is the persistent current in the qubit loop~\cite{F1}, $\Phi_{0}=h/2e$ is the flux quantum. $g=MI_{p}I_{r}/\hbar$ is the qubit-resonator coupling strength. $I_{r}=\sqrt{\hbar\omega_{r}/L_{r}}$ is the zero-point current of the resonator, $L_{r}$ is the resonator inductance.
The qubit frequency $\omega_{qb}=\sqrt{\Delta^2+\varepsilon^2}$ is shown in Fig.~\ref{fig2} by a white dashed line. $\sigma_{z}$ and $\sigma_{x}$ are Pauli matrices. $a^{\dagger}$ ($a$) is the creation (annihilation) operator of the resonator field with  the resonance frequency $\omega_{r}/2\pi$, shown in Fig.~\ref{fig2} as a white horizontal dashed line.

The qubit energy gap $\Delta/2\pi$ is tuned to be equal to the resonator frequency $\omega_{r}/2\pi=2.417$ GHz by applying a long dc bias in the $\alpha$ bias line. Fig.~\ref{fig2}(a) shows the spectroscopic measurement result, an anti-crossing due to the qubit-resonator coupling is clearly observed at the optimal point of the qubit. By fitting the spectrum lines we found the coupling strength $g/2\pi=9.14$ MHz.

To realize switching on/off the qubit-resonator coupling, we apply  a control field with frequency $\omega_0$ to the $\alpha$-loop via the $\alpha$ bias line. This results in a longitudinal interaction Hamiltonian $H_{\rm L}=\hbar\lambda_z\cos(\omega_{z}t) \sigma_{z}$. $\lambda_z$ is determined by the amplitude of the applied rf driving current $I_{\alpha}$ in the $\alpha$ bias line. At the optimal point, i.e., $\varepsilon=0$, the qubit-resonator Hamiltonian Eq.~(\ref{eq:1}) can be reduced to the well-known Jaynes-Cummings Hamiltonian, $H_{\rm JC}=\hbar\Delta\sigma_{z}/2+\hbar\omega_{r} a^{\dagger}a+\hbar g( a^{\dagger}\sigma_{-}+a\sigma_{+})$. By performing a unitary transform $U=\exp[-i2\lambda_{z}\sin(\omega_{z}t)\sigma_{z}/\omega_{z}]$, the total Hamiltonian $H=H_{\rm JC}+H_{\rm L}$ of the qubit-resonator system is reduced to the effective Hamiltonian~\cite{ZLPRA,YXLNJP}
\begin{eqnarray}
H_{\rm eff}= \frac{\hbar}{2}\Delta\sigma_{z}+\hbar\omega_{r} a^{\dagger}a+\hbar g_{\rm eff}(\sigma_{-}a^{\dagger}+\sigma_{+}a),\label{eq:2}
\end{eqnarray}
by neglecting all fast oscillating terms (the detailed derivation can be found in Supplementary Information). Here, $g_{\rm eff} = gJ_{0}(2\lambda_{z}/\omega_{z})$ is the effective coupling strength under longitudinal control, where $J_{0}(x)$ is the zeroth-order Bessel function of the first kind. Equation~(\ref{eq:2}) clearly shows that $g_{\rm eff}$ vanishes when
 $2\lambda_{z}/\omega_{z}$ is a zero point of the Bessel function $J_{0}(x)$ in which the first zero is about $x\approx 2.4$~\cite{ZLPRA,YXLNJP}.
 Namely, at a special amplitude  $\lambda_{z}\approx1.2\,\omega_{z}$, the qubit-resonator coupling is switched off. Moreover, the coupling strength
 $g_{\rm eff}$ can be continuously tuned between two values with  opposite  signs by changing ratio $2\lambda_{z}/\omega_{z}$.
 We define the switch on/off ratio $R$ as the ratio between vacuum Rabi frequencies with and without control field. A detailed simulation
 (see Supplementary Information) shows that at a special amplitude $\lambda_{\rm zoff}$, $R$ drops to below $10^{-5}$ for $\omega_{z}/2\pi=150$ MHz.
 In the following, we will call $\lambda_{\rm zoff}$ the {\itshape switch-off point} and a longitudinal control pulse with such an amplitude a
 {\itshape switch-off pulse}. The fast oscillation terms neglected in Eq.~(\ref{eq:2}) result in a small oscillation on the qubit/resonator state
 when the qubit-resonator coupling is switched off, the amplitude of this oscillation decreases rapidly with increasing $\omega_{z}$.

\begin{figure}[h]
\includegraphics[width=8cm]{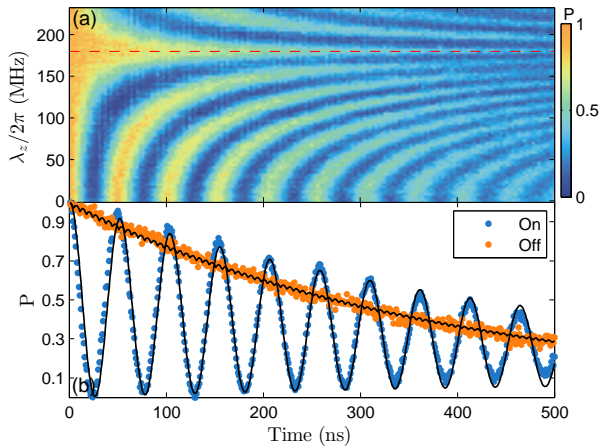}
\caption{{\bf Switching on/off the coherent oscillation between the qubit and the resonator.} (a) Vacuum Rabi oscillations between the qubit and the resonator under different amplitudes $\lambda_{z}$ of longitudinal control pulse. When $\lambda_z$ is increased, the oscillation frequency decreases and reaches a minimum at the {\itshape switch-off point} $\lambda_{z} = \lambda_{\rm zoff}$, indicated by a red dashed line. Color indicates the occupation probability of the qubit excited state. (b) Comparison between the vacuum Rabi oscillation without  (blue) and with (orange) longitudinal control for $\lambda_{z} = \lambda_{\rm zoff}$. When the longitudinal control with the amplitude $\lambda_{\rm zoff}$ is applied to the qubit, the vacuum Rabi oscillation between the qubit and the resonator vanished, indicating that the qubit-resonator coupling is switched off.  Dots are experimental data, black solid curves are theoretical simulation results. P is the occupation probability of the qubit at the excited state.}\label{fig4}
\end{figure}

We first perform spectroscopic measurements on the  qubit-resonator system under different amplitudes $\lambda_{z}$ of longitudinal control fields. As shown in Fig.~\ref{fig2}(b), we found that the amplitude of the anti-crossing gap $\Delta_g = 2g_{\rm eff}$ is decreased to zero and opened again with increasing $\lambda_{z}$. At the {\itshape switch-off point} $\lambda_{\rm zoff}\approx180$ MHz, the amplitude of the anti-crossing $\Delta_g$ becomes undetectable. This is the first evidence that the coupling can be tuned and switched off by the longitudinal control field.

As a further proof of the quantum switch, figure~\ref{fig4}(a) shows the vacuum Rabi oscillations when the amplitude $\lambda_{z}$ of the longitudinal control field is increased from zero. The data clearly shows that the oscillation frequency $\omega_c = 2g_{\rm eff}$ decreases with increasing $\lambda_{z}$. When $\lambda_{z}$ reaches the {\itshape switch-off point} $\lambda_{\rm zoff}$, the oscillation frequency reaches a minimum. Figure~\ref{fig4}(b) shows the comparison of the qubit-resonator dynamics without control pulse (blue curve) and with {\itshape switch-off pulse} (orange curve). At the {\itshape switch-off point} the qubit-resonator dynamics is an exponential decay---the oscillation frequency is too small to be observed on the experimental data, indicating very samll effective coupling $g_{\rm eff}$. Theoretical simulation shows that $g_{\rm eff}/2\pi <100$ Hz, corresponds to the switch on/off ratio $R<10^{-5}$.

\begin{figure*}
\centering
\includegraphics[width=17cm]{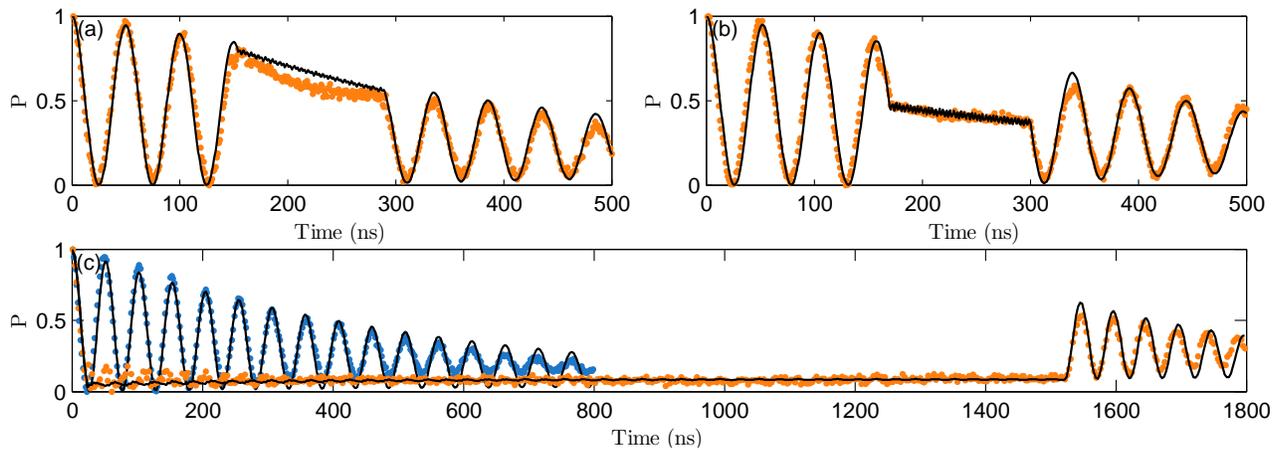}
\caption{{\bf Examples of dynamically switch on/off the qubit-resonator coupling.}  Once a {\itshape switch-off pulse} is applied, the coherent oscillation between the qubit and the resonator is paused and their coupling is switched off. After the {\itshape switch-off pulse},
the coherent oscillation resumes, and the coupling is turned on again.  (a) shows the coupling is switched off and on
when the qubit is in the excited state, (b) shows when the qubit-resonator is in the entangled state.
(c) shows that after the qubit excited state is completely swapped  to the resonator,  the coupling is switched
off for a time duration $1.5\,\mu$s, and then the coupling is switched on,
the coherent oscillation starts again (orange curve). The switch off time is three times longer than the qubit relaxation time $0.45\,\mu$s. As a comparison, the qubit-resonator coherent oscillation
without switching field is also shown as blue dots. All black curves denote theoretical simulations,
dots are experimental data.}\label{fig5}
\end{figure*}

We now demonstrate how to dynamically switch on and off the qubit-resonator coupling. The qubit is initially prepared to its excited state and brought into coherent resonance with the resonator for some time, then a {\itshape switch-off pulse} is applied for a time duration. In Fig.~\ref{fig5}(a), the qubit-resonator coupling is switched off when the qubit is in the excited state and the resonator is in the vacuum. In Fig.~\ref{fig5}(b), the qubit-resonator coupling is switched off when the qubit-resonator is in an entangled state. We found that regardless of what state the system is when the {\itshape switch-off pulse} is applied, the qubit resonator coherent oscillation is paused except for free evolution and decay during the switch-off time interval, after the {\itshape switch-off pulse}, the qubit resonator coherent oscillation resumes. The relaxation times of the qubit and the resonator are $0.45\,\mu$s and $4.6\,\mu$s, respectively. We swap the qubit state into the resonator and switch off the qubit-resonator coupling to store the qubit state in the resonator for a time much longer than the qubit relaxation time. The result is shown in Fig.~\ref{fig5}(c), after a switch-off time of $1.5\,\mu$s, the qubit-resonator oscillation recovers after their coupling is switched on because the system is still the resonator excited state.

In summary, we have demonstrated a highly efficient quantum switch with an rf control field. The coupling strength between a resonator and a qubit can be decreased in magnitude by at least 5 orders. This type of quantum switch can be applied to any qubit system with longitudinal control field and can be easily scaled up since no auxiliary circuit is needed. We believe that this work will also stimulate the research on longitudinal coupling and control between electromagnetic fields and quantum devices, which are normally ignored in traditional quantum optics and atomic physics.

\vspace{1cm}

{\bf Acknowledgements}\; We thank the Laboratory of Microfabrication, Institute of Physics, CAS for the support of the sample fabrication. YXL and DNZ are supported by the National Basic Research Program (973) of China under Grant No.~2014CB921401. XBZ is supported by NSFC under Grants No. 11374344 and
No. 11574380, the 1000 Youth Fellowship program in China and the Chinese Academy of Science.  KN acknowledges support from the Japanese MEXT Grant-in-Aid for Scientific Research on Innovative Areas ¡°Science of hybrid quantum systems¡± (Grant No. 2703). YXL and DNZ are also supported by  NSFC under Grant No. 91321208.

{\bf Author Contributions} YXL and XBZ conceived the study. XBZ and YW designed the experiment, YW, XBZ and YRZ designed the sample, HD, XBZ, ZGY and YJZ fabricated the sample, YW and YRZ carried out the measurements and data analysis with XBZ and YXL providing supervision. YXL developed the theoretical mode. LPY performed theoretical simulations and wrote the supplementary material under the guide of YXL and CPS. DNZ and LL advised on the experiment. WJM and KN provided theoretical support. YXL, YW, XBZ and WJM wrote the manuscript with contributions from all authors. XBZ and YXL  designed and supervised the project.

\clearpage
\section{METHODS}
{\bf The sample.} Our sample is a gap tunable flux qubit inductively coupled to a $\lambda/4$ coplanar resonator. A flux qubit~\cite{F1} is a superconducting loop interrupted by three Josephson junctions, one of them is $\alpha (\alpha<1)$ times smaller then the other two, which are identical and characterized by Josephson energy $E_J$ and charge energy $E_c$. The qubit states correspond to clockwise and anti-clockwise persistent currents $\pm I_{p}$ in the qubit loop, $I_{p}\approx 400$nA in our sample. When the magnetic flux bias in the qubit loop $\Phi_{\epsilon}=\Phi_{0}/2$, called the optimal point, the two persistent current states are degenerate, quantum tunneling lifts this degeneracy, forming an energy gap $\hbar\Delta$. Once $E_J$ and $E_c$ are fixed, the energy gap of a flux qubit is determined by $\alpha$, the ratio of critical current between the smaller junction and the big ones. Since the resonator frequency is fixed, to have the qubit coupled to the resonator at the optimal point where the coherence properties are the best, it is desirable to have the energy gap tunable. In our sample, the smaller junction has been replaced by a dc-SQUID, called the the $\alpha$-loop, making the energy gap tunable in a wide range. The longitudinal control pulse for switching on/off the coupling is also applied to the qubit through the $\alpha$-loop. To achieve high magnetic flux bias stability, we use a gradiometric design for the qubit loop, making the qubit insensitive to global flux fluctuations. The qubit state is detected by a readout dc-SQUID inductively coupled to the qubit. The sample is fabricated on a silicon wafer using electron beam lithography patterning, aluminium evaporation and lift-off techniques, Josephson junctions are formed by using the standard Dolan bridge technique.

{\bf Tuning the energy gap of the qubit.} To obtain long coherence time, the qubit has to work at the optimal point, where the qubit frequency $\omega_{qb}$ is the qubit energy gap $\Delta$. Since the qubit energy gap deviates from the design value in most cases due to limited fabrication precision, to have the qubit coupled to the resonator while working exactly at the optimal point, that is $\Delta = \omega_r$, we first need to tune $\Delta$ to match the resonator frequency $\omega_r$. We apply a long dc pulse to the $\alpha$ bias line to induce a magnetic flux change in the $\alpha$ loop, modulate the value of $\alpha$ thus change the value of the qubit energy gap. Extended Data Figure 1 shows the result, the energy gap can be tuned in a wide range.

{\bf Creation of the longitudinal control pulse.} The longitudinal control pulse is applied by the $\alpha$ bias line. Since the qubit energy gap response to the $\alpha$ bias current is nonlinear, to produce a sine or cosine driving term in $\Delta$, a special waveform has to be used. We fit the $V_{\alpha}-\tilde{\Delta}$ data (Extended Data Figure 1) with a cubic function and got $V_{\alpha}(\tilde{\Delta})=0.2287\tilde{\Delta}^3-2.758\tilde{\Delta}^2+11.14\tilde{\Delta}-15.27$, here $\tilde{\Delta} = 10^{-9}\Delta/2\pi$ is the qubit energy gap in frequency unit GHz. To generate a longitudinal control term $\hbar\lambda_z\sigma_{z}\cos(\omega_z t)$ when the qubit is coupled to the resonator without detunning, the waveform need to be generated is $V = 10^{-9} V_{\alpha}[\omega_{r}+2\lambda_z\cos(\omega_z t)]/2\pi$. This waveform is generated by a Tektronix AWG5014 arbitrary waveform generator and sent to the $\alpha$ bias line through a series of filters and attenuators, the attenuators convert the voltage pulse into current.

\clearpage
\section{Extended Data}
\begin{figure}[h]
\centering
\includegraphics[scale=0.4]{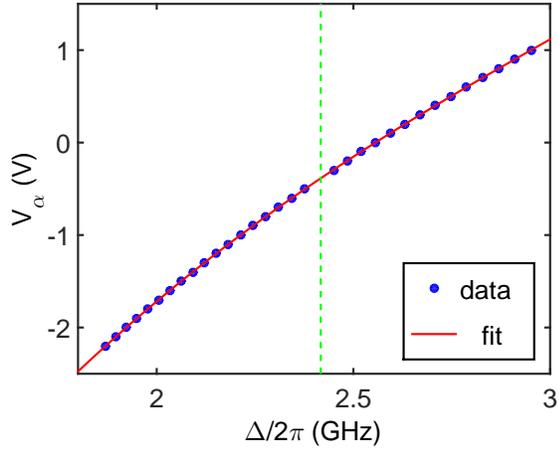}
\caption{{\bf Tunability of the qubit energy gap.} By appling dc bias current in the $\alpha$ bias line, the qubit energy gap $\hbar\Delta$ can be tuned in a wide range. A dc voltage pulse with amplitude $V_{\alpha}$ was generated by a Tektronix AWG5014 arbitrary waveform generator and sent to the $\alpha$ bias line through a series of filters and attenuators, the attenuators convert the voltage pulse into bias current. $V_{\alpha}$ is proportional to the $\alpha$ bias current in the bias line. Dots are experimental data, the red line is a cubic fit of the data. The green dashed line markers out the resonator frequency $\omega_r/2\pi = 2.417$GHz.}\label{FigS1}
\end{figure}

\end{document}